\documentclass[twocolumn]{aastex62}

\usepackage{amssymb,amsmath}
\usepackage[inline]{enumitem} 

\newcommand{\alfven}{Alfv{\'e}n\,}

\received{}
\revised{}
\accepted{}
\submitjournal{ApJ}

\shorttitle{Recurrent CME-like eruptions - Part II}
\shortauthors{Syntelis et al.}

\begin{document}

\title{Recurrent CME-like Eruptions in Emerging Flux Regions. II. Scaling of Energy and Collision of Successive Eruptions}

\correspondingauthor{P. Syntelis}
\email{ps84@st-andrews.ac.uk}

\author{P. Syntelis}
\affiliation{St Andrews University,
Mathematics Institute,
St Andrews KY16 9SS,
UK}

\author{V. Archontis}
\affiliation{St Andrews University,
Mathematics Institute,
St Andrews KY16 9SS,
UK}

\author{K. Tsinganos}
\affiliation{Section of Astrophysics, Astronomy and Mechanics, Department of Physics, University of Athens,\\ Panepistimiopolis, Zografos 15784, Athens, Greece}

\begin{abstract}
We present results of three-dimensional MHD simulations of recurrent eruptions in emerging flux regions. The initial numerical setup is the same with the work by \citet{Syntelis_etal2017} (hereafter, Paper I). 
Here, we perform a parametric study on the magnetic field strength ($B_0$) of the emerging field.
The kinetic energy of the produced ejective eruptions in the emerging flux region ranges from $10^{26}-10^{28}$~erg, reaching up to the energies of small Coronal Mass Ejections (CMEs). The kinetic and magnetic energies of the eruptions scale linearly in a logarithmic plot.
We find that the eruptions are triggered earlier for higher $B_0$ and that $B_0$ is not directly correlated to the frequency of occurrence of the eruptions. 
Using large numerical domains, we show the initial stage of the partial merging of two colliding erupting fields. The partial merging occurs partly by the reconnection between the field lines of the following  and the leading eruption at the interface between them. We also find that tether-cutting reconnection of the field lines of the leading eruption underneath the following eruption magnetically links the two eruptions. Shocks develop inside the leading eruption during the collision.
\end{abstract}


\keywords{Sun: activity -- Sun: interior --
                Sun: Magnetic fields --Magnetohydrodynamics (MHD) --methods: numerical
               }

\section{Introduction}

Most of the activity in the Sun appears to be directly connected with the properties of the solar magnetic fields. Due to dynamo action, magnetic fields, e.g. in the form of flux tubes, are generated inside the convection zone and give rise to sunspots and Active Regions (ARs) when they emerge to the solar surface \citep{Parker_1955}. ARs are the sources of the most intense solar phenomena, such as flares and Coronal Mass Ejections (CMEs). The emergence of magnetic flux (EMF) is also associated with less energetic and smaller-scale events, such as small filament eruptions and micro-sigmoids \citep[e.g.][]{Raouafi_etal2010}. Most eruptive phenomena are commonly related to twisted magnetic flux tubes in the solar atmosphere called flux ropes (FRs). 
A number of observational studies identify the presence of FRs in ARs before an eruption and study their formation \citep[e.g.][]{Green_etal2009, Cheng_etal2011, Green_etal2011,Patsourakos_etal2013}, the pre-eruptive phase \citep[e.g.][]{Canou_Amari2010,Vourlidas_etal2012,Syntelis_etal2016}, and the triggering of the eruptions \citep[e.g.][]{Zuccarello_etal2014,Chintzoglou_etal2015,Yardley_etal2018}.
It is also common for a single AR to produce more than one eruptions \citep[recurrent eruptions, e.g. ][]{Nitta_etal2001,Wang_etal2013}.
The erupting FRs have been also identified in coronagraphic observations \citep[e.g.][]{Vourlidas_etal2013}. Understanding the FRs formation processes and the triggering mechanisms is critical, since these phenomena affect the terrestrial space environment \citep[e.g.][]{Patsourakos_etal2016}.

A number of models of solar eruptions require the presence of a FR prior to the eruption \citep[e.g.][]{Torok_etal2005,Mackay_etal2008,Manchester_etal2008,Torok_etal2011}. Other models demonstrate the formation of a FR in a highly sheared AR during the eruption \citep[e.g.][]{Antiochos_etal1999,Lynch_etal2008}. 
An important parameter for the FR formation prior to the eruption is the presence of shearing and converging motions along and towards a polarity inversion line (PIL) \citep[e.g.][]{Magara_etal2001,Archontis_Torok2008,DeVore_etal2008,Aulanier_etal2010}. Due to these motions, a strong current forms above the PIL. There, the field lines reconnect and start forming a FR \citep[e.g.][]{vanBallegooijen_etal1989}. 
Flux-emergence models have shown that shearing and convergence motions develop naturally during the partial emergence of a sub-photospheric FR \citep[e.g.][]{Manchester_2001, Fan_2001, Manchester_etal2004}. The expansion of the emerging field above the solar surface is commonly associated with untwisting of the embedded field lines and rotation of the emerging bipolar pair of sunspots \citep{Fan_2009,Sturrock_etal2015, Sturrock_etal2016}. Such motions can induce further shearing of the field lines along the PIL. 

How the FR becomes eruptive is still an open issue.
Two ways have been proposed to drive its eruption. One way is through a non-ideal process (e.g. magnetic reconnection) and the other is through ideal MHD instabilities (or catastrophe).

An example of magnetic reconnection that leads to an eruption is the tether-cutting mechanism. There, the magnetic field enveloping the FR reconnects through a current sheet below the FR \citep[e.g. ][]{Moore_etal1980,Moore_etal1992}, commonly referred to as ``flare current sheet''. The upwards release of tension of the reconnected field lines at the ``flare current sheet'' causes fast acceleration of the FR. A variation of this process was found in Paper I. In the latter study, the envelope magnetic field reconnects with low-lying, highly sheared field lines  (J-like loops). 
Another example of magnetic reconnection leading to an eruption is the break-out reconnection between the envelope field and an external magnetic field. If the external field has the appropriate orientation (preferably anti-parallel), it reconnects with the envelope magnetic field \citep[e.g. ][]{Antiochos_etal1999, Karpen_etal2012, Archontis_etal2012, Leake_etal2013}. This reconnection removes the downwards tension of both the envelope and the external field, causing the FR to move upwards and erupt ejectively. On the other hand, if the external field lacks the appropriate orientation  (e.g. it is parallel to the envelope field), then the external field and the magnetic envelope will not reconnect. In such a configuration, a FR eruption would be suppressed by the downwards tension of both the envelope and the overlying field \citep[confined eruption, e.g. ][]{Archontis_etal2012, Leake_etal2014}.

Ideal MHD instabilities can also trigger solar eruptions. One example is the helical kink instability \citep[][]{Anzer_1968,Torok_etal2004}.
This instability occurs when the current (twist) of the FR surpasses a critical value that depends on the configuration of the FR (e.g. cylindrical, toroidal) and the line-tying effect \citep[e.g. ][]{Hood_Priest_1981,Torok_etal2004}. Eruptions triggered by kink-instability can become confined by an external magnetic field \citep{Torok_Kliem_2005}.

Another MHD instability associated with solar eruptions is the torus instability \citep{Bateman_1978,Kliem_etal2006}. If a toroidal current channel with major radius $R$ is placed inside an external field that drops as $B_{ext}\propto R^{-n}$, then the current channel will become unstable when $n$ exceeds a critical value, $n_{crit}$. 
The torus or decay index of the field is defined as $n=- \partial B_{ext} / \partial \ln R$. 
To calculate the decay index in simulations and observations, the usual practice is to estimate the envelope field by calculating the potential magnetic field ($B_p$), and then find the decay index as $n=-z \partial \ln B_p / \partial z$ \citep[e.g.][]{Fan_etal2007, Aulanier_etal2010}.
The critical value of the decay index is affected by a variety of parameters, such as the geometry of the FR, its thickness and whether it expands during the eruption \citep{Demoulin_etal2010,Zuccarello_etal2015}. Numerical Studies have reported values for $n_{crit}$ ranging from one to two \citep{Fan_etal2007,Fan_2010,Demoulin_etal2010,An_Magara_2013,Zuccarello_etal2015}.

CMEs occur frequently and at any given day, 2-20 CMEs can be found between the Sun and a radial distance of 1 AU \citep{Lugaz_etal2017}. When a CME occurs after another one, for instance in a homologous \citep[e.g.][]{Liu_etal2014b,Wang_etal2014} or sympathetic manner \citep[e.g.][]{Schrijver_etal2011,Torok_etal2011}, and the second eruption is faster than the first one, the two ejecta can eventually interact, in what is called a CME-CME interaction \citep[e.g.][]{Gopalswamy_etal2001}.
During such interaction, the CMEs' properties such as their speed, size and expansion rates change. The nature of the collision of two CMEs is complicated, as the eruption can be inelastic, elastic or super-elastic \citep[e.g.][]{Shen_etal2017}. The two colliding CMEs almost always interact through their propagating shocks \citep{Lugaz_etal2015}. They can also interact through magnetic reconnection occurring at the interface between the two colliding flux ropes \citep[e.g.][]{Odstrcil_etal2003,Lugaz_etal2005, Chatterjee_etal2013,Lugaz_etal2013}. The total coalescence of two CMEs is refereed to as CME cannibalism \citep{Gopalswamy_etal2001}. The full coalescence of the two colliding structures occur at different distances from the Sun, as this will depend on magnetic reconnection rates, relative speeds, field orientation, CMEs fluxes etc \citep{Manchester_etal2017}.

In Paper I, we studied the onset mechanism of recurrent eruptions in the context of flux emergence simulations. A buoyantly unstable sub-photospheric horizontal flux tube formed a bipolar region which produced four recurrent eruptions. We found that the combination of torus instability and tether-cutting is the main driving mechanism for these eruptions. 
The kinetic energies of the eruptions were $3\times10^{26}-1.5\times10^{27}$~erg and the magnetic energies were around $1\times10^{28}$~erg. Such energies correspond to small scale eruptions.
A geometrical extrapolation was performed to estimate the physical size of the erupting FRs. 
We found that the size of the eruptions could become comparable to the size of small CMEs. 
Moreover, our results showed that the second eruption occurred soon after and it was faster than the first eruption and, thus, parts of the two erupting fields collided, experiencing a partial merging. In Paper I, we couldn't study this merging further, as the eruptions quickly escaped the numerical domain.

In this paper, we extend our work by varying the magnitude of the magnetic field strength of the sub-photospheric flux tube, $B_0$. The increase of $B_0$ leads to an increase of the magnetic pressure and the expansion of the field into the solar atmosphere, which in turn, leads to an increase of the physical size of the eruptions. To follow the evolution of the eruptions in 3D, we performed the parametric study on a larger numerical domain than the one used in Paper I. In this paper, we study the energy content and the recurrence of the eruptions. We also extend the study of the initial phase of the ``collision'' between two successive eruptions.

In Sec.~\ref{sec:initial_conditions} we describe the initial conditions of our simulations. 
Sec.~\ref{sec:overview} is an overview of the simulations. 
In Sec.~\ref{sec:energies} we discuss the energies and the frequency of the recurrent eruptions. 
In Sec.~\ref{sec:beta_alfven} we briefly discuss the plasma $\beta$ and Mach \alfven speed of the erupting fields.
In Sec.~\ref{sec:cannibalism} we discuss the partial merging of the two first eruptions. 
In Sec.~\ref{sec:conclusions} we summarize and discuss our results.


\section{Numerical Setup}
\label{sec:initial_conditions}


We numerically solve the 3D time-dependent, resistive, compressible MHD equations in Cartesian geometry using Lare3D \citep{Arber_etal2001}. The equations, the resistivity form and the normalization units are shown in Paper I.

The initial condition for the simulation is a horizontal flux tube positioned at $-2.1 \ \mathrm{Mm}$. The axis of the flux tube is oriented along the $y$-direction, so the transverse direction is along $x$ and height is in the $z$-direction.
The flux tube's magnetic field is:
\begin{align}
B_{y} &=B_\mathrm{0} \exp(-r^2/R^2), \\
B_{\phi} &= \alpha r B_{y}
\end{align}
where $R=450$~km is the tube's radius, $r$ the radial distance from the tube axis and $\alpha= 0.4$ ($0.0023$~km$^{-1}$) is a measure of the initial twist of the sub-photospheric field. 
For the parametric study, the magnetic field strength of the initial flux tube takes the values of $B_0$= 8 (2400~G), 10.5 (3150~G), 15 (4500~G), 20 (6000~G).
In each simulation, the flux tube is initially in pressure equilibrium. The flux tube is destabilized by imposing a density deficit along the axis of the flux tube, similar to \citet{Archontis_etal2004}:
\begin{equation}
\Delta \rho = \frac{p_\mathrm{t}(r)}{p(z)} \rho(z) \exp(-y^2/\lambda^2),
\label{eq:deficit}
\end{equation}
where $p$ is the external pressure and  $p_\mathrm{t}$ is the total pressure within the flux tube and $\lambda$ is the length scale of the buoyant part of the flux tube. We use $\lambda=5$ ($0.9$~Mm).

We use a large numerical domain of a 1000$^3$ grid with a physical size of 153$^3$~Mm. We use periodic boundary conditions in the $y$ direction and open boundary conditions in the $x$ direction and at the top of the numerical domain. Closed boundary conditions are assumed at the bottom of the numerical domain.

The numerical domain consists of an adiabatically stratified sub-photosheric layer at $-7.2\ \mathrm{Mm}\le z < 0 \ \mathrm{Mm}$, an isothermal photospheric-chromospheric layer at $0 \ \mathrm{Mm} \le z < 1.8 \ \mathrm{Mm} $, a transition region  at $1.8 \ \mathrm{Mm} \le z < 3.2 \ \mathrm{Mm}$ and an isothermal corona at $3.2 \ \mathrm{Mm} \le z < 145.8 \ \ \mathrm{Mm}$.
The stratification layers and resolution are the same as in Paper I.
The initial field-free atmosphere is in hydrostatic equilibrium.


\begin{table}[]
\centering
\caption{Summary of numerical experiments}
\label{tab:energies}
\begin{tabular}{lcccc}
\hline
B$_0$   	&	$E_{mag}$				& $E_{kin}$ (erg)                      \\ \hline
20	  	 	&	1-3$\times10^{29}$		& 1-3$\times10^{28}$                    \\
15   		&	4-6$\times10^{28}$		& 1.5-7$\times10^{27}$                  \\
10.5*   	&	2.5-5$\times10^{28}$	& 5$\times10^{26}$-2$\times10^{27}$     \\
8     		&	2.6-3.6$\times10^{27}$	&  8$\times10^{25}$-1.5$\times10^{26}$  \\ \hline
\end{tabular}
\tablecomments{
Magnitude of initial flux tube's field strength (column 1), range of magnetic and kinetic energies of eruption(s) (column 2 and 3).
The asterisk marks the simulation with the same $B_0$ as in paper I.}
\end{table}

\section{Parametric study: Varying the magnetic field strength $B_0$}
\label{sec:parametric}

\subsection{Overview}
\label{sec:overview}

In Paper I, we used a flux tube with $B_0=10.5$ ($3150$~G) inside a smaller numerical domain ($417^3$ grid points). This is a relatively low magnetic field strength for $z=-2.1$~Mm. The cases with $B_0=15,20$ ($4500$, $6000$~G) have field strengths that could be more indicative for magnetic fields close to the solar surface \citep[e.g.][]{Cheung_etal2007}. A case of lower field strength, $B_0= 8$ ($2400$~G) was also studied.

Flux tubes with higher $B_0$ rise faster inside the solar interior (magnetic buoyancy is $\propto B_0^2$), and carry lower plasma $\beta$. Therefore, the buoyancy instability  criterion, for emergence above the solar surface, is triggered earlier \citep{Acheson1979, Archontis_etal2004}.
Above the solar surface, the flux tubes with higher $B_0$ expand faster due to their larger magnetic pressure and, consequently, they occupy more physical space. For all $B_0$ cases, we find a series of recurrent eruptions following the marked expansion of the field. 

In the $B_0=15, 20$ cases, the expansion of the field is so pronounced that the large numerical domain used is necessary in order to follow the formation of the erupting field before the overall magnetic system reaches the boundaries of the domain.
In the $B=8,10.5$ cases, the expansion of the field is not so pronounced. In these cases, we use a larger numerical domain to i) compare the energies between different $B_0$ and ii) to track the erupting fields further upwards, before they exit the numerical domain. 

\subsection{Energies and recurrence}
\label{sec:energies}

\begin{figure}
\centering
\includegraphics[width=\columnwidth]{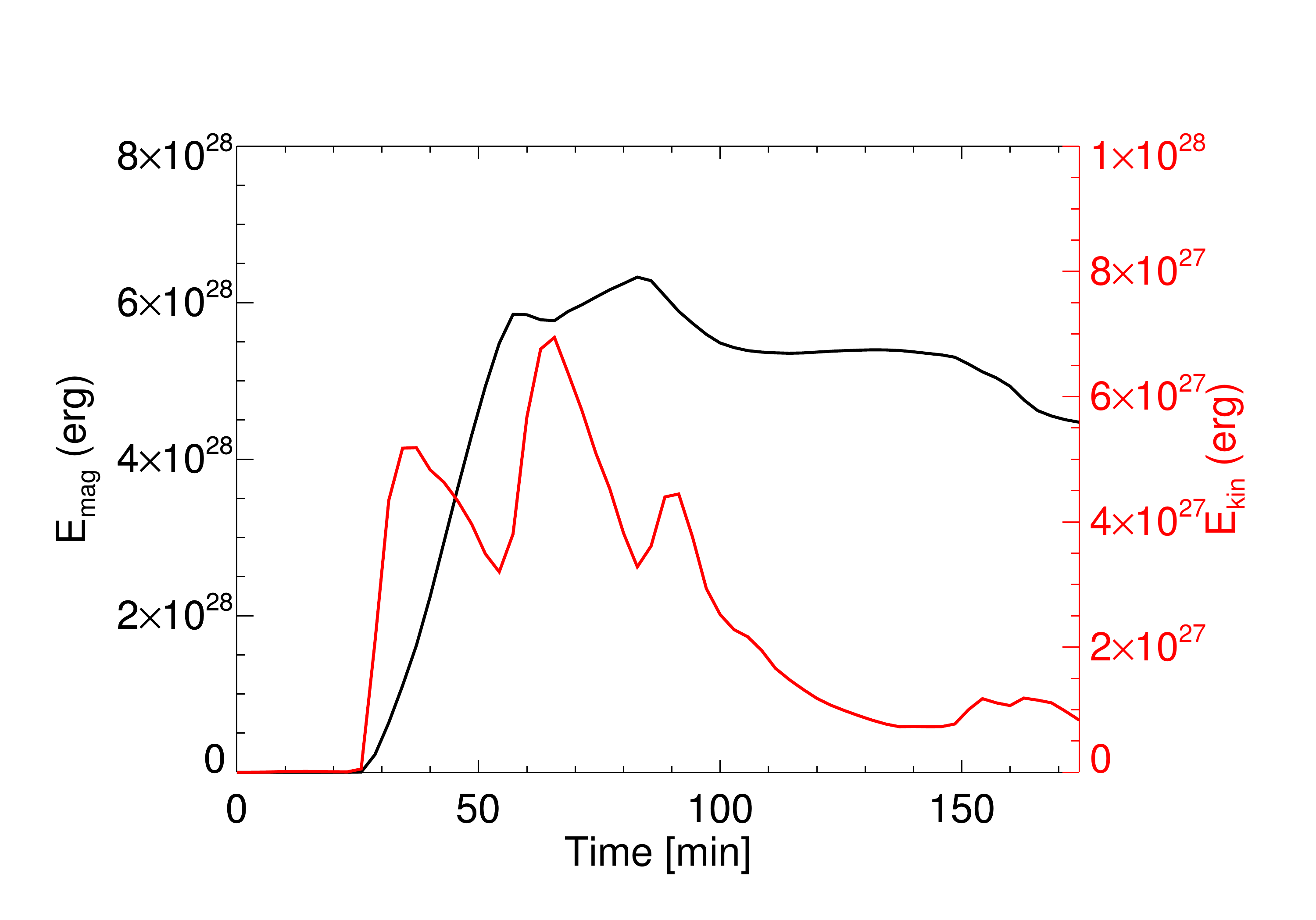}
\caption{ 
Magnetic (black) and kinetic (red) energy above the middle of the photospheric layer ($z=$1.37~Mm) for $B_0=15$. 
}
\label{fig:energy15}
\end{figure}

\begin{figure}
\centering
\includegraphics[width=\columnwidth]{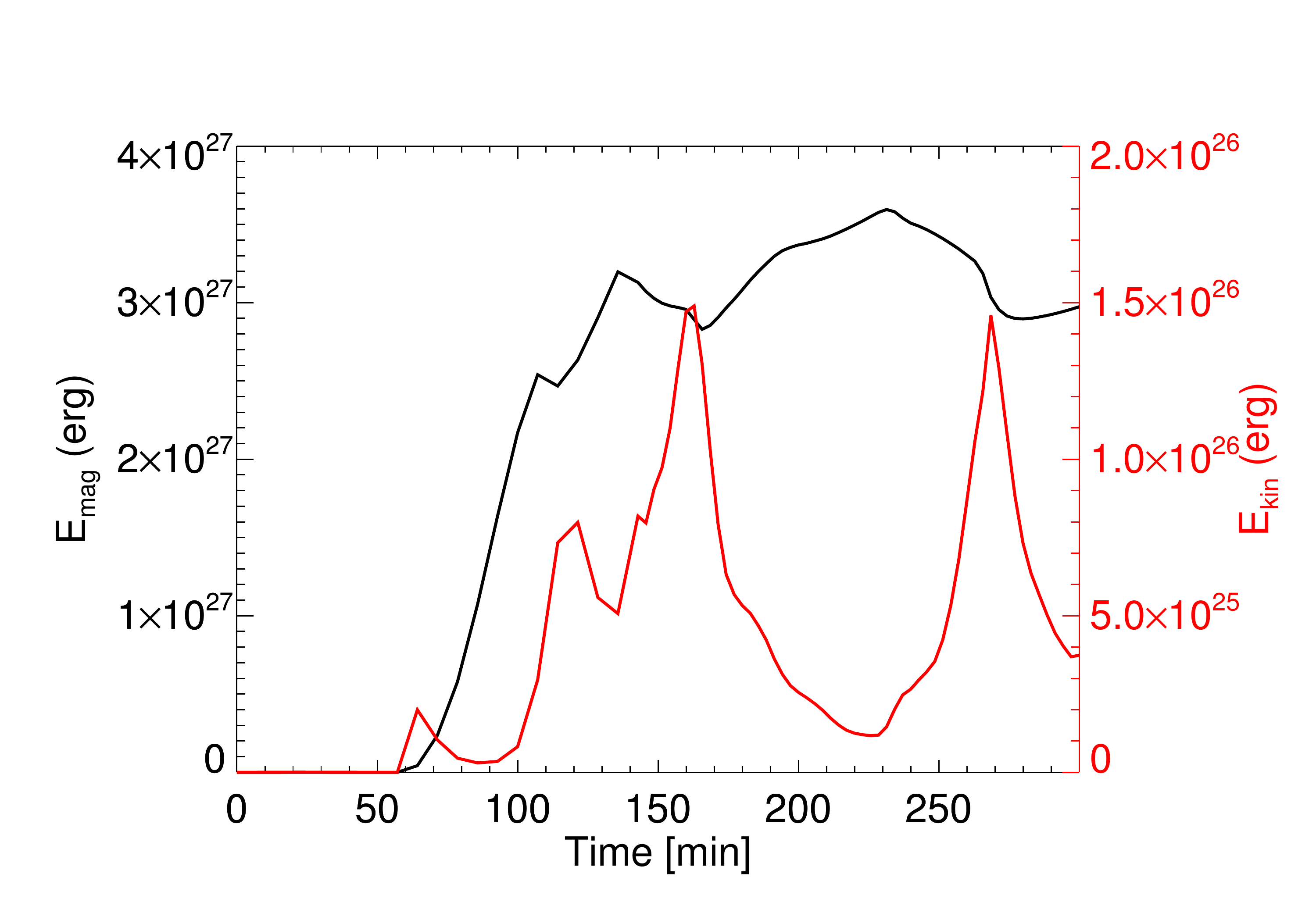}
\caption{ 
Magnetic (black) and kinetic (red) energy above the middle of the photospheric layer ($z=$1.37~Mm) for $B_0=8$. 
}
\label{fig:energy8}
\end{figure}

\begin{figure}
\centering
\includegraphics[width=\columnwidth]{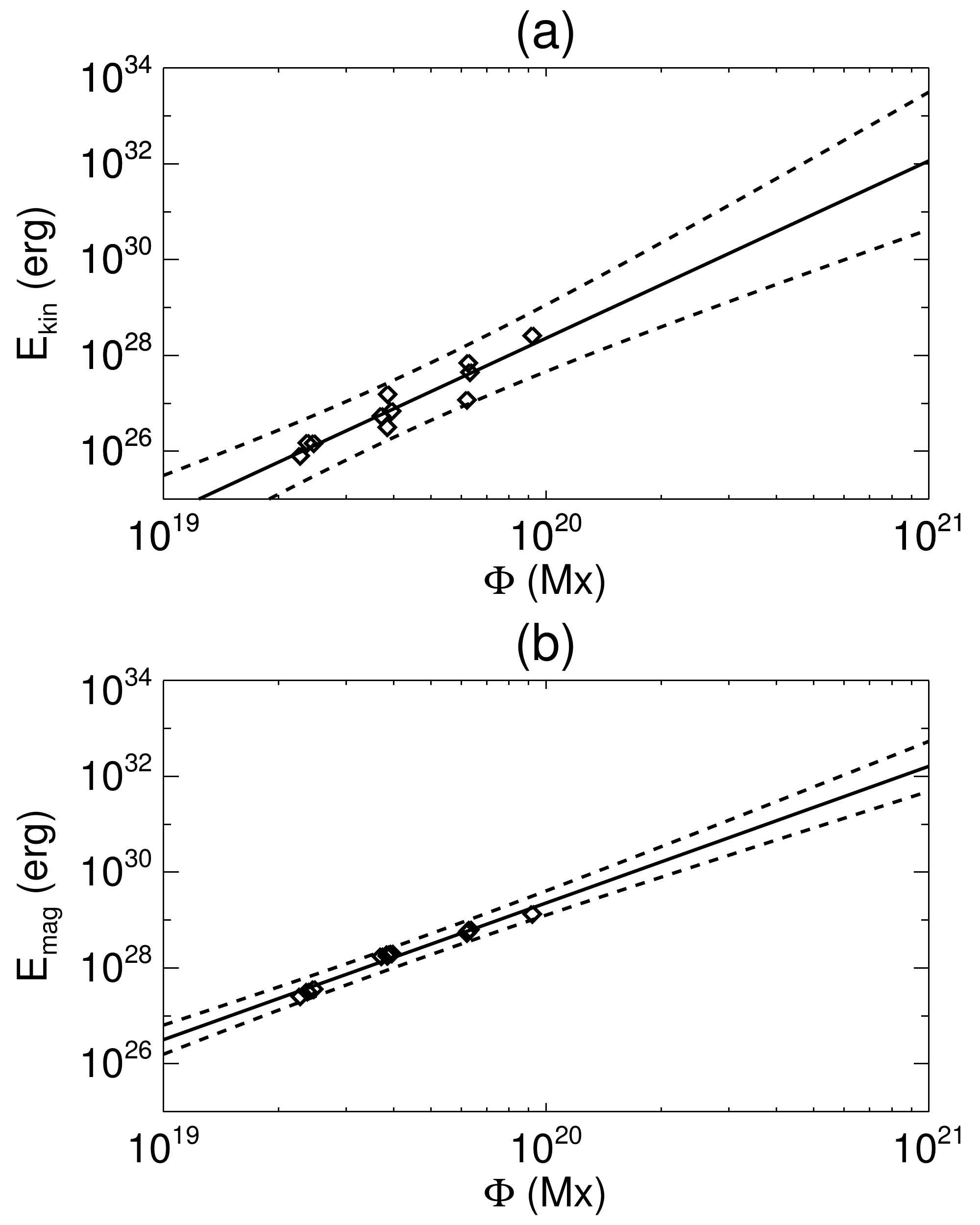}
\caption{ 
Kinetic (a) and magnetic (b) energy maxima over the photospheric density at the time eruptions (diamonds). Energies are measured above the middle of the photospheric layer ($z=$1.37~Mm). Solid lines are linear fitted lines. Dashed lines are the 95\% confidence level of the fitted lines. 
}
\label{fig:scaling}
\end{figure}

\begin{figure*}
\centering
\includegraphics[width=\textwidth]{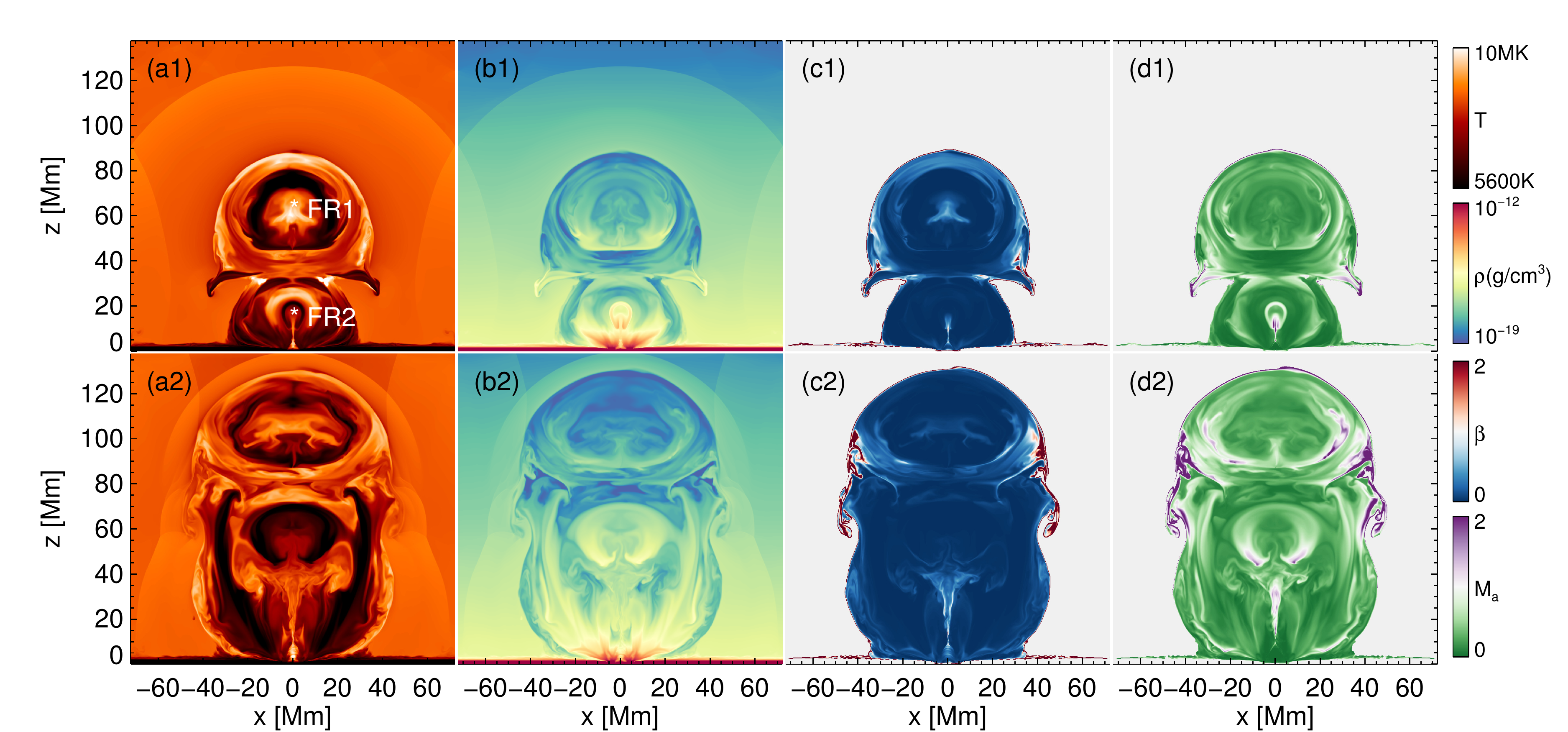}
\caption{
Temperature (first column), density (second column), plasma $\beta$ (third column), and Mach \alfven number (fourth column) measured at the $xz$-midplane for the $B_0=10.5$ simulation at $t=82.8$~min (first row) and $t=88.6$~min (second row).
Asterisks mark the location of the FR center.
The two eruptions ``collide''  between the two times instances.
Grey background in the third and fourth column show the non-magnetized background.
}
\label{fig:eruption_10}
\end{figure*}

Recurrent ejective eruptions are formed in all of the studied cases. The eruption mechanism is very similar to the one described in Paper I (Sec.~3) (i.e. a combination of torus instability and tether-cutting reconnection).

We calculate the kinetic and magnetic energies above mid-photosphere. 
For each $B_0$, we find a series of local maxima that correspond to the eruptive events. 
Figure~\ref{fig:energy15} shows the time evolution of the kinetic and magnetic energy for $B_0=15$. The three kinetic energy maxima associated with eruptions are found at $t$=$65.7,\, 91.4,\, 154.3$~min (the first kinetic energy peak is the initial emergence above the photosphere). The time between the eruptions is $\Delta t=25.7,\,62.9$~min. 
For the $B_{0}=20$ we studied only the first two eruptions. The eruptions are found at $t$=$51.4,\, 128.6$~min with a time delay of $\Delta t=77.2$~min.
For $B_0=8$, the peaks of the kinetic energy (Figure~\ref{fig:energy8}) are found at $t=121.4,\, 162.8,\, 268.5$~min. The time delay between the eruptions is $\Delta t=41.4,\,  105.7$~min. 
In paper I ($B_0=10.5$), the four eruptions were found at $t=74.3,85.7,117.1,197.1$~sec, with $\Delta t=11.4,\, 31.4,\, 80$~min. These are similar to the times of the eruptions using the large domain.
From the above, we find that the first eruption occurs earlier for higher $B_0$.
However, we don't find any direct correlation between the time delay between the eruptions and $B_0$.

The range of the values of the kinetic and magnetic energy maxima for each $B_0$ is shown in Table~\ref{tab:energies}. For $B_0=10.5$, the magnetic energy range of the eruptions is higher than the one reported in Paper I. This is because the energies here are calculated in the large numerical domain. During and after each eruption, the large domain is filled with more magnetic field in comparison to the small domain used in Paper I. Therefore, the large domain's corona contains up to 3 times more magnetic energy after each eruption.

Next, for all of our eruption, we take the values of the kinetic energy peaks and the values of the preceding magnetic energy peaks and plot them as a function of the photospheric flux at that time (diamonds in Figure~\ref{fig:scaling}a and b). We perform a linear regression to asses any potential linear scaling. 
In panel (a), the fitted line (solid line) is 
$\log E_{kin} = (3.7 \pm 0.4) \log \Phi + (-45.7 \pm8.2)$
with an $r^2=0.90$, and in panel (b) is 
$\log E_{mag} = (2.8 \pm 0.4) \log \Phi + (-27.7 \pm 3.4)$
with an $r^2=0.98$.
The 95\% confidence interval of our fitted lines is shown with dashed lines.
The total magnetic field energy above the photosphere correlates well with the photospheric flux.
The kinetic energy associated with the eruptions is more spread as the same emerging region can produce eruptions of different kinetic energies.
As our statistical sample is relatively small ($n=11$), the confidence interval is relatively wide for larger values of the photospheric fluxes. However, based on our extrapolation, for a typical active region flux of $10^{21}$~Mx, we predict energies of $4\times10^{30}-3\times10^{33}$~erg, which are typical energies of CMEs. 

\subsection{Plasma $\beta$ and \alfven speed}
\label{sec:beta_alfven}

Figure~\ref{fig:eruption_10} shows the temperature (first column), density (second column) at the $xz$-midplane of the first two eruption of the $B=10.5$ case. 
In panel (a1), the magnetized plasma between $z\approx35-85$~Mm belongs to the the first eruption. 
The FR center is located inside its central region ($z$=60~Mm, FR1 asterisk). 
Below $z\approx30$~Mm we find the second erupting structure (the FR2 asterisk marks the center of the second FR). 
At that time, the second FR has just entered the fast-rise phase during its eruption. 
Panel (a2) shows a later time when both FRs  undergo a full ejective eruptive phase.
Both erupting structures have a hot and dense central region (panels (b1), (b2)), which is surrounded by a hot and dense outermost region (similar to Paper I). 

In panels (c1) and (c2) we plot the plasma beta, $\beta=P_g/P_m$, where $P_g$ and $P_m$ are the gas and magnetic pressure respectively. Both eruptions have $\beta<1$ everywhere and therefore the eruptions are magnetically dominated.
One exception are the ``flanks'' of the erupting structures (i.e. the interface between the magnetized eruptions and the non-magnetized atmosphere, for e.g. around $x=\pm45$~Mm), where $\beta>1$ (red and white color).
We find that the aerodynamical drag deforms these high beta regions (e.g. around $x=\pm45$~Mm and $z\in[60,100]$~Mm).

\begin{figure*}
\centering
\includegraphics[width=\textwidth]{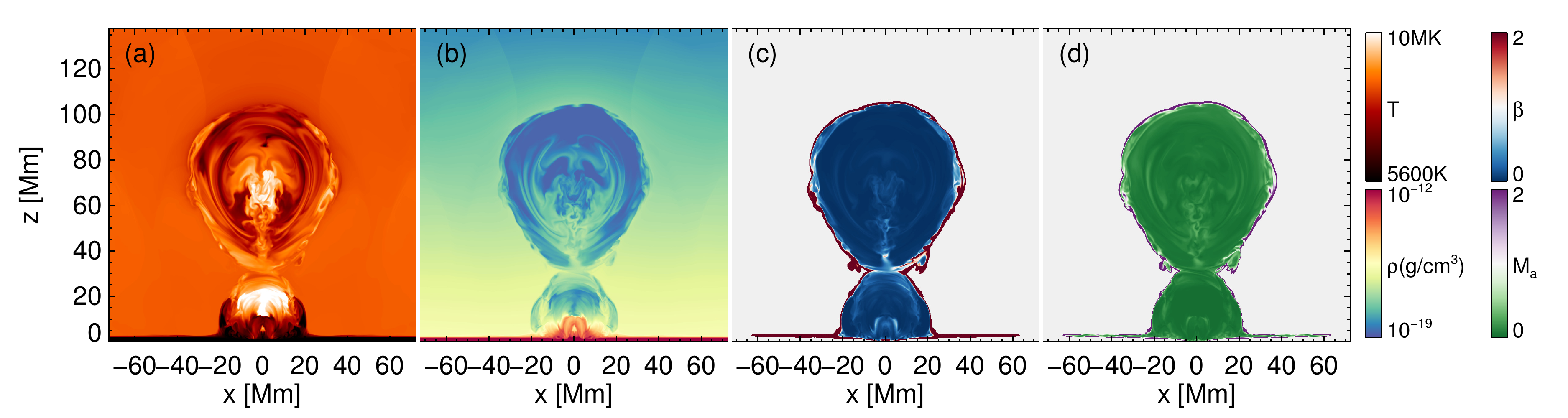}
\caption{
Temperature (a), density (b), plasma $\beta$ (b), and Mach \alfven number (d) measured at the $xz$-midplane for the first eruption of the $B_0=8$ simulation at $t=128.6$~min.
Grey background in the (a) and (b) column show the non-magnetized background.
}
\label{fig:eruption_8}
\end{figure*}

Another region where $\beta$ can be around unity is the flare current sheet. Panel (c2) shows as an example the flare current sheet of the second eruption (white color around $x\in[-5,5]$ and $z\in[5,40]$). 
There, plasma $\beta$ increases locally due to both the increase of the plasma density and temperature and due to the lower magnetic pressure inside the current sheet. 
It has been reported that during fast reconnection high-$\beta$ plasmoids can result from the fragmentation of the current sheet \citep[e.g.][]{Karlicky_etal2012}.

In panels (d1) and (d2) we show the Mach \alfven number of the two eruptions ($M_a = v/v_a$, where $v$ is the magnitude of the velocity field and $v_a = B/ \sqrt{4\pi\rho}$ is the \alfven speed). Inside the erupting structures and around the center of the FR, we find regions of high $M_a$ (purple regions). The high $M_a$ indicates shocks inside the erupting structures. These shocks are formed during the eruption of each of the FRs. 
These shocks will be further discussed in Sec.~\ref{sec:cannibalism}.

The $B_0=15,20$ cases are similar to the $B_0=10.5$ one.
For lower the magnetic field strength case, $B_0=8$, we show the first eruption (
Figure ~\ref{fig:eruption_8}). Here, we do not find regions of $M_a>1$ inside the erupting structure, besides inside the flare current sheet. We still find $\beta\geq1$ regions around the ``flanks'' of the erupting field.

\begin{figure*}
\centering
\includegraphics[width=0.8\textwidth]{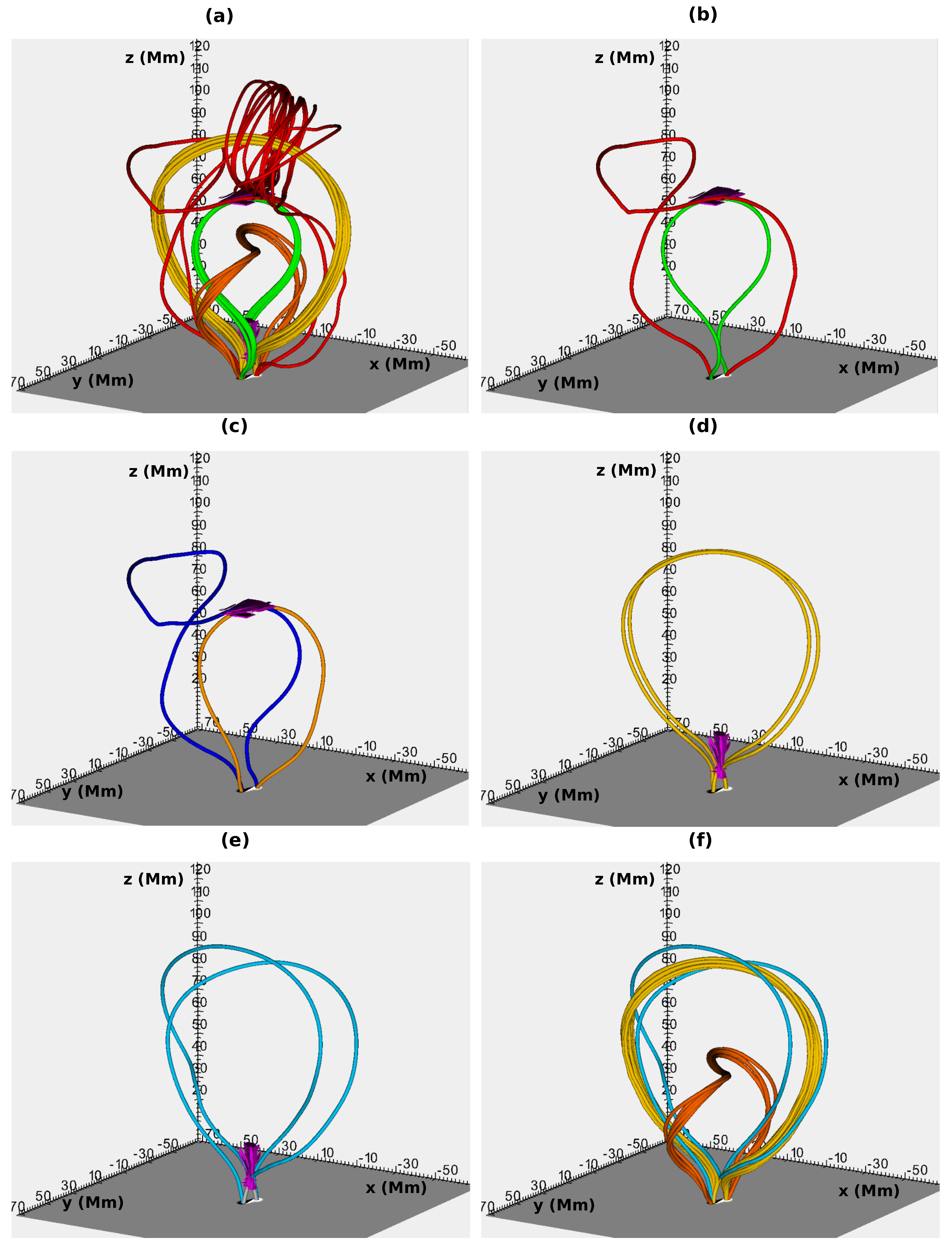}
\caption{
Selected fieldline structures during the partial merging of the first two eruptions, for $B_0$=10.5. 
(a) Yellow (orange) field lines are part of FR1 (FR2). The red field lines surround FR1. 
The green field lines envelope FR2. 
The purple isosurfaces is $\left|J/B\right|>0.1$.
(b) Example red and green field lines before and (c) after reconnection at the current sheet at the inteface between the two ``colliding'' structures.
(d) Example yellow field lines before and (e) after tether-cutting reconnection at the flare current sheet.
(f) The field lines resulted in panel (e) twist around both FR1 and FR2, merging the two structures.
}
\label{fig:cannibalism}
\end{figure*}

\section{Partial merging of the erupting fields}
\label{sec:cannibalism}

In Paper I, we studied the height-time and velocity profiles of the eruptions of the $B_0=10.5$ case. The second erupting FR was faster than the first. Before the second erupting structure exited the numerical domain, it encountered the magnetic field of the first one. This caused a deceleration of the second FR (see second FR's velocity profile, black line, Figure~9a, Paper I). This is what we refer to as ``collision'' of the two erupting magnetic structures. In this paper, we follow this ``collision'' further, by following the eruption further upwards in the large numerical domain.

Figure~\ref{fig:eruption_10} (a1) shows the temperature the two erupting FRs. 
Between the two erupting structures, (around $z\approx33$~Mm and $x=0$~Mm), there is a thin interface with hot plasma. This is part of the flare current sheet underneath the first eruption. 
In Figure~\ref{fig:eruption_10} (a2), when both FR1 and FR2 undergo a full ejective eruptive phase, the first erupting structure is found between $z\approx80-130$~Mm and the second below $z\approx80$~Mm. The two structures ``collide'' at around $z\approx80$~Mm.

Notice that between panel (a1) and panel (a2), the width of leading eruption (its extent along $z$ decreases. This ``contraction'' of the leading eruption is commonly found during CME-CME interactions and it is due to the compression of the leading structure during the ``collission'' \citep[e.g.][]{Manchester_etal2017}.
Also, during the ``collision'' of the two erupting structures, the high $M_a$ regions inside the first eruption (panel (d1)) become more enhanced and extended (panel (d2)).
This demonstrates that the compression between the colliding structures can form/enhance shocks inside two colliding eruptions. Such shocks inside CMEs have been reported to be formed/enhanced during the collision of CMEs/ICMEs \citep[e.g.][]{Lugaz_etal2015, Lugaz_etal2017}.

To understand better the structure of the erupting field(s), we use visualization of selected field lines, which are traced from various locations within the overall erupting system. In Figure~\ref{fig:cannibalism} (a), we trace field lines from the FR1 center in yellow (as in Figure~7 (c) of Paper I). 
With red color, we trace field lines from an area that surrounds the central region of the erupting field (i.e. the hot circular cross-section area, located around $-40$~Mm$ < x < 40$~Mm and $85$~Mm$< z < 125$~Mm in Figure~\ref{fig:eruption_10} (a2)).  
These field lines had undergone Envelope-Envelope tether-cutting reconnection, as discussed in Paper I (and are similar to the red field lines in Figure~7 (c) of Paper I).
The field lines, which go through the center of the FR2 are shown in orange color. The erupting field that envelopes FR2 is plotted with green lines. 
These lines are traced from the high temperature region around $z\approx75$~Mm in Figure~\ref{fig:eruption_10} (a2). The orange and green field lines correspond to the FR and envelope field in Figure~8 (f) of Paper I (with yellow and green color, respectively).

Studying the topology of the field lines, we notice that the two initially separated structures start to partially merge after they ``collide''. This partial merging happens through magnetic reconnection of their fieldines, which occurs mainly at two different sites.
The first reconnection site is shown in Figure~\ref{fig:cannibalism} (b) (this is the regions around $z\approx80$~Mm, Figure~\ref{fig:eruption_10}, (a2)). 
A current sheet (purple isosurface) is formed between the green and the red field lines. These sets of field lines reconnect with each other at this current sheet. This reconnection forms field lines that start from the one end of the green envelope field lines, twist around FR1, and end at the footpoints of the red field lines (blue line, Figure~\ref{fig:cannibalism} (c)). 
The reconnection forms also a field line that starts from the one end of the green envelope field lines and end at the footpoints of the red field lines (orange line). This new line has less downwards magnetic magnetic tension than the green line of Figure~\ref{fig:cannibalism} (b). Therefore, after the reconnection it is possible that the upwards rise of FR2 will be further assisted.

The second reconnection site is shown in Figure~\ref{fig:cannibalism} (d) (purple isosurface). This is the location of the flare current sheet of the second eruption. 
As FR1 (yellow lines) move upwards, its footpoints move towards the vicinity of the flare current sheet of the second eruption. 
There, the yellow FR1 field lines reconnect with other FR1 field lines via tether-cutting  (i.e. the envelope-envelope tether-cutting mechanism discussed in Paper I). 
This reconnection forms twisted field lines (e.g. light blue line, Figure~\ref{fig:cannibalism} (e)) and adds flux to the low lying post-reconnection arcade (e.g. grey line).
Notice that the light blue line twists around both FR1 and FR2, linking the two structures (Figure~\ref{fig:cannibalism} (f)).
Further more, the upwards tension release of the field lines similar to the light blue field line push will push FR2 upwards, bringing FR1 and FR2 further closer together.
This reconnection can occur many times. For instance, a field line similar to the light blue one of panel (e) can reconnect with lines similar to the yellow lines of panel (d), forming fieldlines with high twist.

Because the system is very complex, it is very likely that there are more reconnection sites within the overall erupting volume of the field.
However, the blue and light blue field lines are the two main sets of new field lines found after the ``collision'' of the two magnetic structures. For simplicity, we refer to the linkage of the two erupting field structures, through the blue and light blue lines, as partial merging. 

We were unable to study the partial merging in the other $B_0$ cases. The reason for that is that in all other cases, the time interval between the onset of two successive eruptions is larger than the one found in the $B_0=10.5$ case. 
Therefore, even if the second eruption was faster than the first eruption, the first eruption would have partially or fully exited the numerical domain before the second erupting structure ``collided'' with it.

\section{Summary and Discussion}
\label{sec:conclusions}

In this work, we performed a parametric study on the magnetic field strength ($B_0$) of a subphotospheric magnetic flux tube. We focused on the evolution of the eruptions, which occurred after the emergence of the magnetic field at the solar surface and the formation of a small AR. In this study, we used a relatively large numerical domain (1000$^3$ grid points, $145.6^3$~Mm). The values of $B_0$ used are shown in Table~\ref{tab:energies}. In all the $B_0$ cases, we found that recurrent eruptions were triggered above the PIL of the ARs. 
The $B_0$=8, 10.5 cases ($B_0=10.5$ is also the paper I case) yielded kinetic energies of $8\times10^{25}-2\times10^{27}$~erg. Such energies correspond to small scale eruptive events.
For $B_0=15,\,20$, the kinetic energy of the eruptions ranged from $1\times10^{27}-3\times10^{28}$~erg, which is comparable to the kinetic energy of small scale CMEs \citep{Vourlidas_etal2010}. 
A further increase of $B_0$ (e.g. up to $B_0=35$) might produce energies of the order of an average CME (around 10$^{29}$~erg). Such a work would require even larger numerical domains, which are computationally very expensive. Also, a spherical grid would be required to study in detail the eruptions of that scale. 

In all of our cases, the first eruption is formed earlier in time for higher $B_0$. 
For $B_0=8$, the frequency of the eruptions is lower compared to the $B \ge 10.5$ and above. For the higher $B_0$, we do not find a clear correlation between the frequency of the eruptions and the increase of the initial sub-photospheric magnetic field strength.
Therefore, we cannot associate the frequency of the eruptions with $B_0$.

For all of our eruptions (11 in total) we plotted their kinetic energy and the atmospheric magnetic energy prior to the eruption as a function of the photospheric flux. We found that both quantities follow linear scaling in a log-log plot. Our sample ranges across a relatively small range of photospheric fluxes, and the resulting energies of the eruptions are comparable to small-scale prominence eruptions \citep[mini or micro CMEs e.g.][]{Innes_etal2010b,Raouafi_etal2010,Hong_etal2011} and small CMEs   \citep[e.g.][]{Vourlidas_etal2010, Reeves_etal2015}. Similar energies are reported in simulations of small scale eruption such as the solar jets \citep[e.g.][]{Raouafi_etal2016,Liu_etal2016}.
The extrapolation of our linear fit to active region fluxes ($10^{21}$~Mx) results to typical kinetic energies for large scale CMEs ($10^{30}-10^{33}$~erg). If such a scaling is indeed occurring in solar eruptions, it would connect eruptions of both smaller and larger scales.
It would be interesting to study whether such a linear scaling remained after including events with typical energies of small and large CMEs and events with typical energies of jets (both standard and blowout ones). Such a scaling would be a strong indication on the jet-CME connection \citep[e.g.][]{Wyper_etal2017}.

We also studied the distribution of plasma $\beta$ inside the erupting structures. We found that indeed the eruptions are magnetically driven since $\beta<1$ inside the erupting field. Plasma $\beta$ becomes greater than one at the outermost region of the magnetized structures (the ``flanks'' of the eruptions). This high beta region can be deformed by the aerodynamical drag. 

In Paper I, we studied in detail the $B_0$=10.5 case using a small numerical domain (417$^3$ grid points, $64.8^3$~Mm). There, the second eruption was found to be faster than the first one, with signs of ``collision'' between the two successive eruptions. In this work, using the larger numerical domain, we were able to follow this ``collision'' further upwards.  We found that it leads to the partial merging of the two magnetic structures. 
The merging occurred by the reconnection of field lines, at two (at least) sites.
The first reconnection site was located at the interface between the two colliding structures adding twist around the first erupting FR while removing flux from the following erupting FR.
The second reconnection site was the flare current sheet underneath the second erupting FR. There, the field lines of the leading FR reconnected through tether-cutting, forming lines that surrounding both FRs, magnetically linking the the two structures.

Simulations studying CME-CME interaction of two erupting FRs propagating through the solar wind \citep[e.g.][]{Odstrcil_etal2003,Lugaz_etal2005,Lugaz_etal2013}, show that reconnection occurs always at the interface between the first and second eruption. 
Depending on the relative orientation of the FR's axis, this process can lead to the full or not merging of the two structures \citep{Lugaz_etal2013}.
Here, we show that the tether-cutting reconnection of the leading FR's field lines can magnetically connect the two erupting structures with an additional way that has not been reported before.
We cannot conclude whether the two FRs of our simulation would fully merge and become co-spatial \citep*[as in the study of ][]{Chatterjee_etal2013}. It is likely that our experiment describes aspects of the initial phase of a cannibalistic process.

During the collision of the two structures, regions of high \alfven Mach number develop. We interpret these structures as shocks inside the colliding FRs. Observations of CME-CME interactions in the interplanetary medium have reported the development of shocks inside the colliding CMEs \citep[e.g.][]{Lugaz_etal2015, Lugaz_etal2017, Shen_etal2017}.

\acknowledgments
The Authors would like to thank the Referee for the constructive comments.
This project has received funding from the Science and Technology Facilities Council (UK) through the consolidated grant ST/N000609/1. 
The authors acknowledge support by the Royal Society.
This work was supported by computational time granted from the Greek Research \& Technology Network (GRNET) in the National HPC facility - ARIS. 

\bibliographystyle{aasjournal}
\bibliography{bibliography}


\end{document}